\documentclass[reprint,aip,jap]{revtex4-1}
\usepackage[T1]{fontenc}% pour les lettres accentuées de la biblio
\usepackage{graphicx}
\usepackage{comment}
\usepackage{amsmath}
\usepackage{amssymb}
\usepackage{hyperref}
\usepackage{xspace}
\newcommand{\tern}{Cu$_{60}$Ti$_{20}$Zr$_{20}$\xspace}
\newcommand{\SWMD}{$\phi_{\text{SW}}$-MD\xspace}
\newcommand\QE{\emph{QE}\xspace}
\newcommand\rdf{\emph{rdf}\xspace}
\newcommand\rdfs{\rdf{}s\xspace}
\begin{document}

\title{Structure of the glass-forming metallic liquids by ab-initio and classical molecular dynamics, a case study: quenching the \tern alloy
%Quenching a glass-forming metallic liquid, a case study by ab-initio and classical molecular dynamics: the ternary \text{\tern} alloy. \\
%Structure of glass-forming metallic alloys by ab-initio and classical molecular dynamics, a case study: quenching the liquid \text{\tern} ternary alloy..
}
\author{S. Amokrane, A. Ayadim and L. Levrel }
\affiliation{Groupe ``Physique des Liquides et Milieux Complexes'', Facult\'{e} des Sciences et
Technologie, Universit\'{e} Paris-Est (Cr\'{e}teil), 61 av. du G\'{e}n\'{e}ral de Gaulle,
94010 Cr\'{e}teil Cedex, France}
\date{19/06/2015}

\begin{abstract}
  We consider the question of the amorphization of metallic alloys by melt
  quenching, as predicted by molecular dynamics simulations with
  semi-empirical potentials. The parametrization of the potentials is
  discussed on the example of the ternary Cu-Ti-Zr transition metals alloy,
  using as reference the ab-initio simulation. The pair structure in the
  amorphous state is computed from a potential of the Stillinger Weber form.
  The transferability of the parameters during the quench is investigated
  using two parametrizations: from solid state data, as usual, and from a new
  parametrization on the liquid structure. When the adjustment is made on the
  pair structure of the liquid, a satisfactory transferability is found
  between the pure components and their alloys. The liquid structure predicted
  in this way agrees well with experiment, in contrast with the one obtained
  using the adjustment on the solid. The final structure, after quenches down
  to the amorphous state, determined with the new set of parameters is shown
  to be very close to the ab-initio one, the latter being in excellent
  agreement with recent X-rays diffraction experiments. The corresponding
  critical temperature of the glass transition is estimated from the behavior
  of the heat capacity. Discussion of the consistency between the structures
  predicted using semi-empirical potentials and ab-initio simulation, and
  comparison of different experimental data underlines the question of the
  dependence of the final structure on the thermodynamic path followed to
  reach the amorphous state.
\end{abstract}
\maketitle

\section{Introduction}

Since their discovery in the late eighties \cite{Inoue1988}, multicomponent
bulk metallic glasses (BMG) have been the subject of numerous experimental and
theoretical studies. The unique physical, mechanical, and corrosion properties
of these novel materials indeed enable a variety of applications
\cite{Trexler,Gong,Niinomi}. A particular attention has been paid to the BMGs
based on late transition metals, especially in the Cu-based family
\cite{Inoue2,D-Xu}, including alloys with Zr \cite{Jayanta,Venkataraman}, Ti
\cite{Ma}, or Al \cite{Cu-Zr-Al}. These studies have shown that the glass
forming ability (GFA) of an alloy and its properties in the amorphous state
depend on several factors such as the nature of its components, its
composition, or the cooling rate. Simulations are then helpful for a
systematic exploration of the parameters space, including domains that are
difficult to study experimentally. They are also useful for understanding the
underlying physics at an atomic scale. Most often, they consist in classical
molecular dynamics (MD) with semi-empirical atomic potentials, such as the
embedded atom model (EAM) \cite{EAM}, the tight binding, second moment
approximation (TB-SMA) \cite{TB-SMA}, and Finnis-Sinclair (F-S) potentials
\cite{F-S} and their variants. Their parameters are adjusted from available
structural and thermodynamic data. EAM potentials for fcc elements are given
for example in Ref. \onlinecite{Sheng} and n-body potentials for ternary
alloys are reviewed in Ref. \onlinecite{Lia}. For the Cu family see Ref.
\onlinecite{Cu-Zr-Ag} for Cu-Zr-Ag, or Ref. \onlinecite{Mendelev} for Cu-Zr
alloys. For the latter, the parameters were determined from a combination of
ab-initio calculations and experimental data. Other methods, such as the force
matching technique in which ab-initio potential energies are used as input,
have also been used (see for example Refs. \onlinecite{FM1} for the Cu-Zr-Al
alloy and \onlinecite{FM2} for Cu-Zr). Carefully adjusted potentials can then
be used for practical applications\cite{Sheng}, like searching optimized
compositions \cite{Lia}.

The central question in this classical route remains the transferability of
the parametrized force fields, to different state points or compositions, for
example. The transferability, especially for the n-body potentials has been
discussed e.g. in Ref. \onlinecite{Alemany} for fcc metals in the liquid phase
(TB-SMA), Ref. \onlinecite{Noya} for Ni-Al (EAM), or Ref. \onlinecite{Ozdemir}
for liquid Pd-Ni alloys (modified Sutton-Chen \cite{S-C} potential).
Furthermore, the data used for the fits -- say the cohesive energy, lattice
constants, etc. -- are usually relative to the solid phase. It is then not
obvious that the same parameters will also describe the properties of liquid
and amorphous metals \cite{Alemany}, which lack the periodicity which is
important for the electronic structure of the solid \cite{Foiles}. In related
areas, this question is discussed in Refs. \onlinecite{Madden,Tafipolsky}, or
\onlinecite{Harvey} for the use of this approach in thermodynamic integration
methods.

As an alternative, one may consider first-principles simulations for which
the question of transferability does not arise. Their computational cost
however makes them often unfeasible without resorting to supercomputers. This
holds even with non all-electron ones, which use pseudopotentials and
approximate functionals. This includes paths involving a wide range of
variation of the parameters, such as in cooling rate studies \cite{Cheng2} --
see also Refs. \onlinecite{Cluster-Glue1,Cluster-Glue2}. They are thus used
occasionally, say for supplementing the information required to fit the
phenomenological potentials
\cite{Tafipolsky,Zope,abinitconstr,Neural_Net,Eshet}. To this end, when the
amorphous state is obtained by quenching a liquid, namely through a path
between two disordered states, it may be preferable to adjust the parameters
on the liquid properties.

Since there is no systematic means to devise the optimum transferability (for
alternatives see for example Ref. \onlinecite{Tafipolsky}), new elements
collected in representative cases are useful to ascertain this question. In
this work, we shall illustrate this on the example of the ternary \tern alloy
which has been studied recently by experiment (see references in Refs.
\onlinecite{Inoue2,Jiang,Dai,Dai2,Ze-xiu,Mattern2,Cu-Ti-Zr_Durisin}) and
simulation with parametrized potentials \cite{Teichler1,Qin,Dalgic,Fuji} for
its importance as a model system of transition-metal-based BMGs, and its
technological relevance. To our knowledge, this alloy has not been studied
from first-principles simulation, in contrast with Cu-based binary alloys (see
for example Refs. \onlinecite{Jakse1,Jakse2} for Cu-Zr). Using molecular
dynamics at the Born-Oppenheimer level (BOMD) and density functional theory
(DFT)\cite{DFT}, as implemented in the Quantum Espresso (\QE) package
\cite{QuantumEspresso}, we present here such a study intended to highlight
some questions raised by the use of locally adjusted atomic potentials in
multicomponent glass forming alloys (a recent review of ab-initio molecular
dynamics (AIMD) methods applied to glass formation can be found in Ref.
\onlinecite{Pasturel}). This will be done by comparing the pair structure
determined from the ab-initio and classical routes, in the liquid and at
ambient temperature which is well below the critical temperature of the glass
transition of this alloy. By adjustment to the high-temperature structure of
the alloy determined from AIMD, a new set of parameters of the effective
potential is determined. Their transferability is investigated by comparison
with the ab-initio structure at the final temperature. The corresponding
critical temperature of the glass transition is estimated from the behavior of
the heat capacity. The stability of the structure predicted using the
semi-empirical potential with respect to the AIMD simulation is discussed.
Comparison with other simulations and available experiments is made to
underline the importance of thermodynamic path followed to reach the amorphous
state. % revoir

This paper is thus organized as follows: in section II, we detail the
methodology we use for this purpose. In section III, we present the main
results concerning the ab-initio ($g^{ab}_{ij}(r)$) and classical ($g_{ij}(r)$)
radial distribution functions (\rdfs) and the estimated glass transition
temperature. The system size dependence is finally discussed. We conclude this
paper by a summary of the main conclusions.

\section{Methodology}

\subsection{General method}

The question of transferability is actually independent of the reality of the
reference data used to parametrize the potentials. As we could not find
experimental data for the pair structure of \tern at high temperature,
we decided to use as ``experimental'' data the \rdfs $g^{ab}_{ij}(r)$ of the
alloy obtained from AIMD in the liquid state. We
postpone to the next section the question of its relationship with the actual
structure of the liquid alloy. The ``classical'' \rdfs $g_{ij}(r)$ are determined
by MD simulation using the simplified (i.e. without 3-body terms)
Stillinger-Weber \cite{SW} (SW) potential used by Han and
Teichler \cite{Teichler1}:
\begin{equation}
  \phi_{SW}(r)=A\left(\frac{1}{(\alpha r)^n}-1\right)\exp\left(\frac{1}{\alpha r-a_1}\right), \;\; 0<r \leqslant \frac{a_1}{\alpha}.
\end{equation}
These simulations will be referred to as \SWMD. The set of parameters of the
SW potential adjusted using data from the solid \cite{Teichler1} will be
designated as $\{a^s_i\}$ and those from the liquid, determined as $\{a^l_i\}$. The parameters $\{a^l_i\}$ are varied until the pair structure determined with the
\SWMD simulation ($g_{ij}(r)$) agrees reasonably well with the ab-initio one
($g^{ab}_{ij}(r)$). Once this is achieved, the alloy is cooled down to
$T=300$~K.  Starting from the last equilibrated MD configuration so obtained, the AIMD is
run at ambient temperature, until satisfactory equilibration is reached. This
is followed by a series of accumulation steps for obtaining the statistical
averages. A first comparison can thus be made between the two final sets of
\rdfs at $T=300$~K, $g^{ab}_{ij}(r)$ and $g_{ij}(r)$. As a test of the role of
the initial configuration, we perform an ``instantaneous'' quench by using as
input at $T=300$~K a configuration equilibrated in (or near) the liquid state.
One may then compare the \rdf determined with finite and infinite cooling
rates (with necessary caution in the latter case). We also used the last
configuration obtained with the $\{a^s_i\}$ parametrization. Before presenting
these comparisons, we give below some technical details of the classical and ab-initio runs.

\subsection{Computational details}
\subsubsection{Classical MD}
The extensive \SWMD runs were performed using the LAMMPS package \cite{LAMMPS}
in the isobaric NPT ensemble using the Nose-Hoover integration, with a time step
$dt = 0.0025$ ps. Over a total length of about $6\, 10^6 $ steps,  the last
$2 \,10^6 $  were used to compute the averages.  All potentials were cut at
$r=  a/\alpha $ and finite size effects were investigated by changing the
particle number from 260 to 1372. The quenches were performed in the NPT
ensemble (at $p=0$), either with a stepwise change of temperature or a
continuous one. Different cooling rates were considered (besides an
instantaneous quench): $3\,10^{10}$~K\,s$^{-1}$ and a 10 times slower one.   

\subsubsection{Ab-initio MD}
The AIMD simulations were run using of the plane-wave self consistent field
(PWscf) code in the \QE package with no modification other than adding the
computation of the six \rdfs $g^{ab}_{ij}(r)$ at regular time intervals (say
every twelve hours) to monitor equilibration and production steps directly on
the pair structure. The ions dynamics used a time step of $dt=30$~a.u.
($1.45\, 10^{-3}$~ps), the equations of motion being integrated with the Verlet
algorithm. Between $N=128$ (pure components) and $N=240-260$ (ternary alloy)
particle numbers were considered,  and a simple velocity rescaling method was
used to fix the temperature. (N,V,T) simulations were performed, but, to
bracket the desired zero pressure, the volume was initially varied starting
from the value for an ideal mixture (see below). A series of  short runs
(about  200-300 steps) with different box sizes were made to monitor the
fluctuation of the pressure above and below $p=0$. Full NPT simulations using
the variable cell-shape method implemented in the \QE distribution indeed
proved too costly on the rather small 24-core machine we used. 

The accuracy of the AIMD simulation depends mostly on the quality of the
electronic energy computed in the scf cycles. The quality of the
exchange-correlation functional used in the DFT treatment of the electronic
structure and some parameters such as the number of plane waves used in the
expansion of the Kohn-Sham orbitals are known to be important, but they were
found not to be critical here (i.e. for the determination of the pair
structure). After several trials, an energy cutoff of 30~Ry and a density
cutoff of 240~Ry, in the range of the recommended values, were found sufficient
to ensure a satisfactory convergence. Ultra-soft (USP) \cite{USP}
pseudopotentials from the \QE library \footnote{Cu.pbe-n-van\_ak.UPF,
Ti.pbe-sp-van\_ak.UPF, Zr.pbe-nsp-van.UPF from the \QE library. Details on
their generation are given in the file headers and accompanying text files.},
all with the PBE functional of Perdew \emph{et al.} \cite{PBE} were used (with
$Z=11, 12, 12$ valence electrons for Cu,Ti and Zr). While more accurate
\cite{Kresse} projector augmented-wave (PAW) \cite{PAW} pseudopotentials are
available, our tests did not indicate a significant effect at the level of the
\rdfs (see below the discussion for pure Cu). The convergence threshold for the
scf cycles was $10^{-6}$~Ry for typical energies of $3\, 10^4$~Ry. Due to the
large supercell size used (typically 15~\AA), calculations were performed at
the $\Gamma$ point only (except for the density of states discussed in section
\ref{sectionDOS}). In these conditions and using standard values (for metals)
for the other settings of the PW code, one scf cycle for a total number of 2964
electrons (for 260 atoms) takes on our machine roughly 2300~s.

\section{Results}
\subsection{High temperature structure}
\subsubsection{Liquid copper}
\begin{figure}[htbp]
\centering
\includegraphics[clip,width=7.0cm]{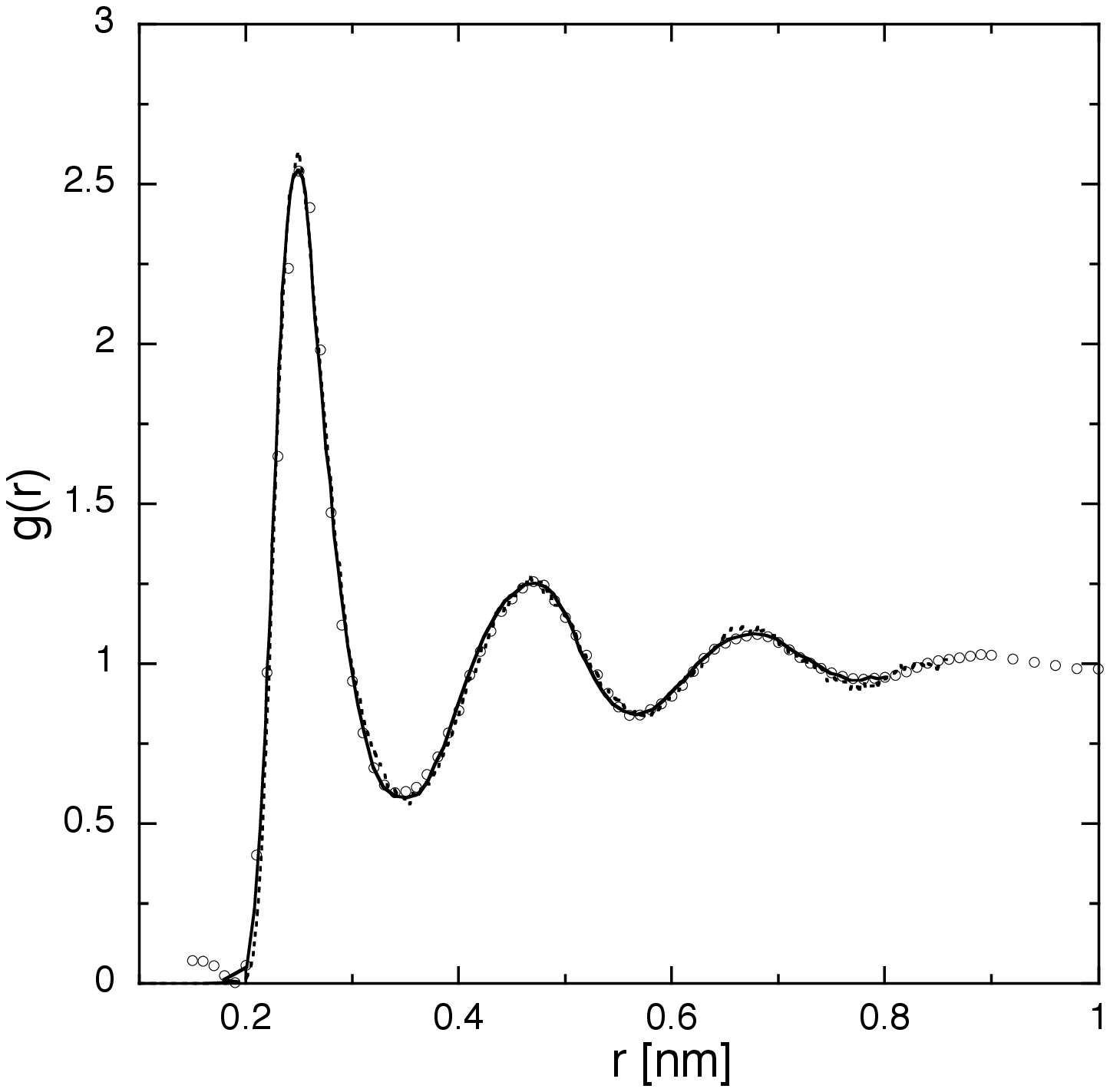}
\caption{Radial distribution function of liquid Cu at $T=1623$~K.   \\Symbols:
  experimental X-ray diffraction data \cite{Waseda} interpolated to 1623~K.
  Full line: simulation of Ref. \onlinecite{Widom} (PAW). Dots: this work (USP).}
  \label{figureCupur}
\end{figure}

To test various parameters in the AIMD runs, we computed the \rdf
$g^{ab}_{Cu}(r)$ for pure Cu. $N=128$ particles were placed in a cubic box
with a supercell size $L=23.3$~a.u. to achieve the experimental equilibrium
density $\rho = 0.074$~\AA$^{-3}$ at atmospheric pressure \cite{Waseda}.
$g^{ab}_{Cu}(r)$ was computed in the liquid at $T=1623$~K to allow comparison
with the AIMD of Ganesh and Widom \cite{Widom} (PAW pseudopotentials) and the
experimental data of Ref. \onlinecite{Waseda}. The AIMD parameters are those
indicated above. The result is shown in figure (\ref{figureCupur}).

To test the sensitivity to the pseudopotential, both USP and PAW ones were
used. The value of the pressure is actually sensitive to the pseudopotential,
but the difference between USP and PAW pseudopotentials (a few kilobars here)
has a negligible effect on the equilibrium density and hence on the \rdf. With
the USP, the peak is higher by two percents but the oscillations are of
similar quality as those with the PAW potentials. The similar agreement
between USP and PAW calculations with experiment led us to use the USP since
the scf cycles are then less time-consuming. Computing an accurate \rdf for Cu
has actually no novelty per se since this can be achieved using a much faster
parametrized version of the generalized pseudopotential theory for the noble
metals, as known since the mid-eighties \cite{Christ}. This calculation was
only intended to check various parameters in the scf cycles, besides giving an
indication of the expected structure in the alloy. A previous study by Jakse
and Pasturel \cite{Jakse1,Jakse2} indeed showed a relative insensitivity of
the ab-initio Cu-Cu \rdf (at least in the first peak) to copper
concentration in the binary Cu$_x$Zr$_{1-x}$ alloys. We could not find a
similar study for Cu-Ti but the data in Refs. \onlinecite{CuTi_Ristic} and
\onlinecite{CuTi_Dalgic2} do not indicate qualitative effects of alloying with
Ti. The important point then is that since Cu is the majority component in
\tern, the \rdf of Cu should be similar in the alloy and in the pure
component, which is very well described by AIMD. This is also a good
indication that the structure determined from the simulation of a small
periodic system is representative of the one in the true alloy, as discussed
later.

\subsubsection{Liquid alloy}
\begin{figure}[htbp]
\centering
\includegraphics[clip,width=7.0cm]{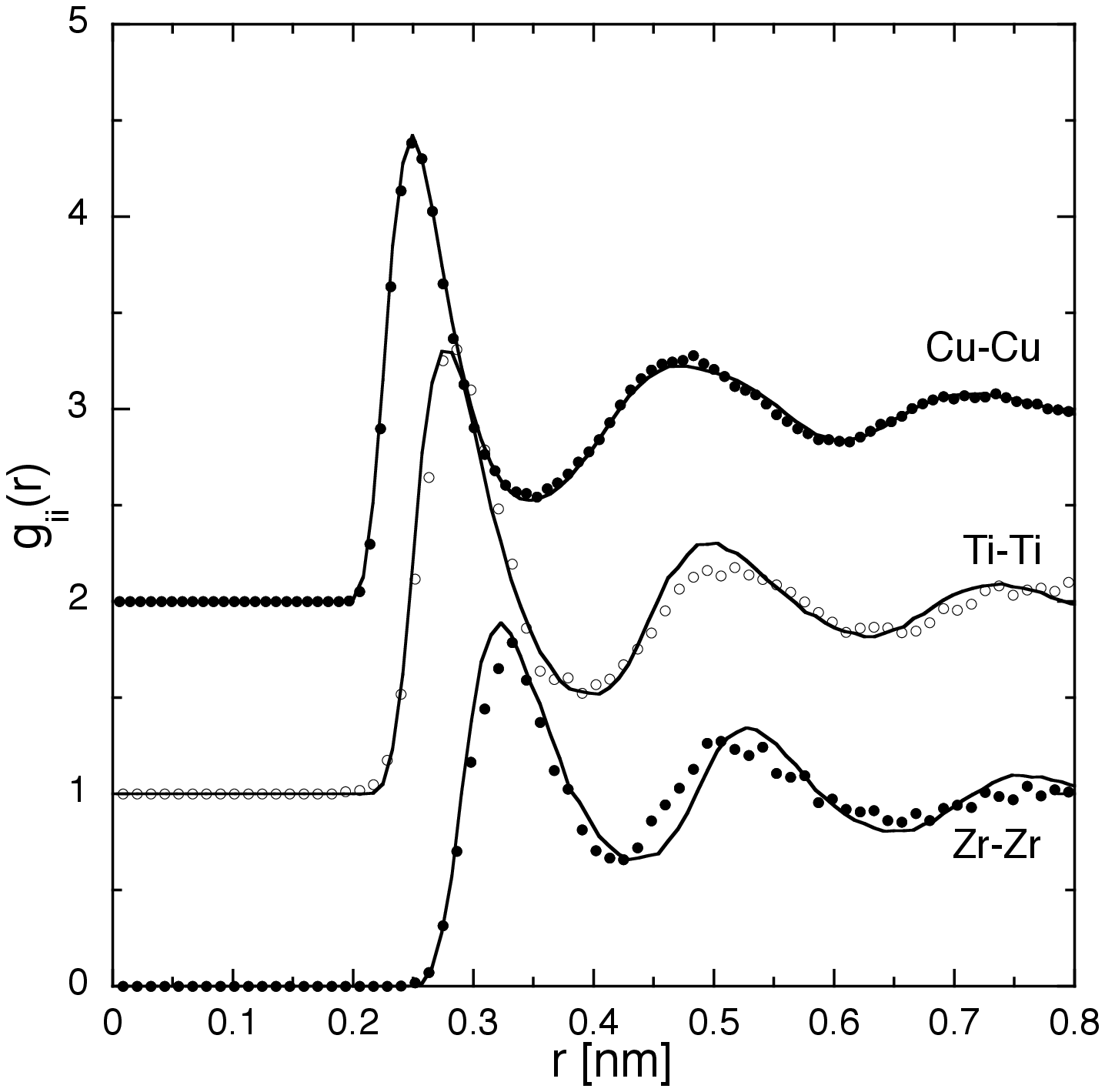}
\includegraphics[clip,width=7.0cm]{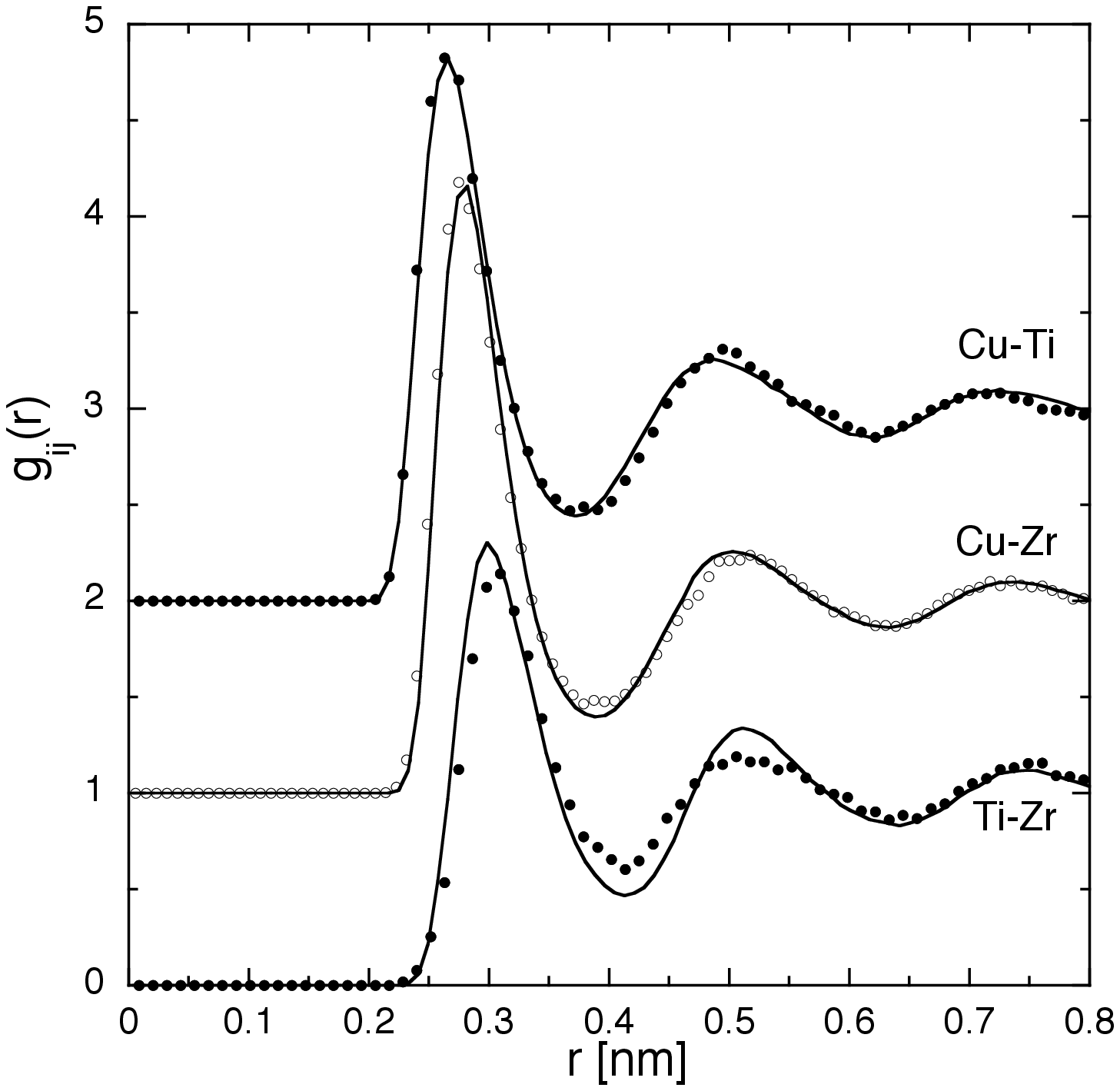}
\caption{Radial distributions functions in the \tern alloy at $T=1600$~K.
  \\ Symbols: AIMD; lines: \SWMD (with parameters $\{a^l_i\}$ as in table
  I).}
\label{gij1600K}
\end{figure}

The partial \rdfs of \tern at 1600~K determined from the AIMD with ultra-soft
pseudopotentials are shown in figure (\ref{gij1600K}). They were obtained with
$N=260$ particles ($156+52+52$ atoms for Cu, Ti, and Zr respectively) to keep
the CPU time in a tolerable range. The system was initially prepared as
follows: an initial density was computed from the known experimental densities
of the pure liquid components, assuming an ideal mixture behavior. This gives
$\rho\sim 0.062$~\AA$^{-3}$. For the corresponding volume, a cubic box is
filled with the $N$ particles randomly distributed on the vertices of a cfc
lattice (plus a small random displacement), to form the supercell. Its size is
then varied about the initial value $L\sim 30.5$~a.u. to bracket the zero
pressure at 1600~K. Within a tolerance of a few kilobars, $L$ converged rather
rapidly, the final value being $L=30.914$~a.u. The corresponding volume was
then used to continue the AIMD runs in the NVT condition. After about 500 steps
to monitor equilibration, 1500 steps of accumulation generate, in about 45 days
of computation, reasonably smooth curves for $g_{Cu-Cu}$, as well as for the
cross ones involving Cu, somewhat less so for the minority species Ti and Zr .
Note for the latter a significantly larger effective diameter and a
``smoother'' $g(r)$, clearly calling for a refined parametrization, possibly
with the three-body terms (with the present model, we recently  reached a total
of 2475 steps which showed no significant effect aside from a progressive
smoothing of the data).  As a check of equilibrium, we noted that the total
energy is very stable, with a very small relative fluctuation and the pressure
recorded during the last 839 steps fluctuates about a slightly negative but
well-defined average value\footnote{See supplemental material at
\texttt{[energy\_1600K.pdf]} for a plot of the energy and at
\texttt{[pressure\_300\_1600.pdf]} for a plot of the pressure.}.

The set of parameters $\{a^l_i\}$ was then determined by (hand) adjustment on
these ab-initio \rdfs, in the liquid ternary alloy. They are given in table I.
 \begin{center}
\begin{table}[htdp]
\begin{tabular}{cccccccc}
\hline\hline
      & $A$ (eV)       && $\alpha^{-1}$ (\AA) && $a_1$    & $n$    \\ \hline
Cu-Cu & 1.135\,(0.485) && 2.290\,(2.275) && 1.681\,(1.681) & 9\,(9) \\
Cu-Ti & 1.725\,(1.695) && 2.325\,(2.300) && 1.805\,(1.794) & 7\,(7) \\
Ti-Ti & 1.364\,(1.588) && 2.380\,(2.350) && 1.960\,(2.056) & 9\,(4) \\
Cu-Zr & 1.942\,(1.943) && 2.450\,(2.496) && 1.792\,(1.792) & 8\,(8) \\
Ti-Zr & 1.959\,(2.722) && 2.530\,(2.481) && 1.900\,(1.968) & 7\,(3) \\
Zr-Zr & 1.695\,(3.655) && 2.710\,(2.646) && 1.950\,(1.855) & 9\,(3) \\
\hline\hline
\end{tabular}
\caption{Parameters $\{a^l_i\}$ of the SW potential adjusted on the ab-initio
  liquid structure. The values between parentheses shown for comparison are
  the parameters $\{a^s_i\}$ adjusted on the solid, from Table II of Ref.
  \onlinecite{Teichler1}.}
\end{table}
\end{center}

The corresponding $g_{ij}$ are drawn as lines in figure (\ref{gij1600K}).

It is stressed that this set of parameters is by no means unique. The goal was
not to determine the best parametrization but to investigate whether one that
gives a reasonably accurate initial structure (in comparison to a given
reference) keeps its quality after a quench down to the amorphous state. For
this reason, the much longer runs that would have been needed to improve the
statistics for the minority species were not attempted. With this caveat in
mind, it is clear from the table that the values determined from the
adjustment to the liquid structure differ considerably from those of Ref.
\onlinecite{Teichler1}. We note in particular the large difference for the exponent parameters which affect the short-range behavior of such effective  potentials (see above the remark for Zr).  This is a clear evidence of non-transferability from
the solid to the liquid phase.

\subsubsection{Transferability of the parametrization $\{a^l_i\}$ at high temperature}
\begin{figure}[htbp]
\centering
\includegraphics[clip,width=7.0cm]{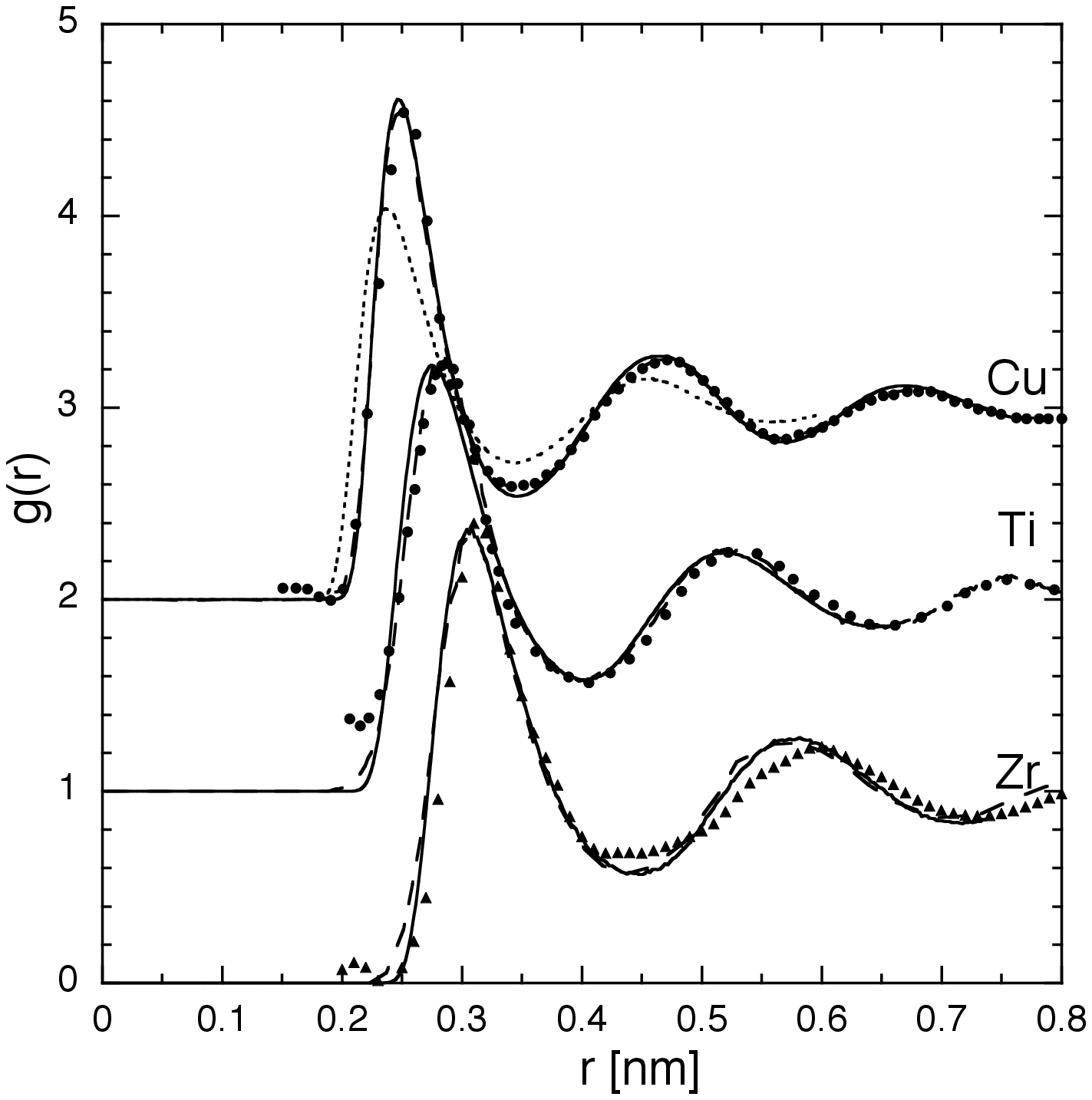}
\caption{Radial distribution functions of liquid Cu (1623 K), Ti (1973 K) and Zr (2173 K).\\
  Symbols: experiment (Cu and Zr: Ref. \onlinecite{Waseda}, Ti: Ref.
  \onlinecite{Ti-exp}); full curves: \SWMD, $\{a^l_i\}$; dotted curve: \SWMD,
  $\{a^s_i\}$ for Cu; dashed lines: AIMD (from Ref. \onlinecite{Jakse3} for Zr
  at 2500 K).}
\label{figureCuTiZrPurs}
\end{figure}

The transferability of the parameters $\{a^l_i\}$ in the liquid state was
first checked on the pure components. We show in figure
(\ref{figureCuTiZrPurs}) the \rdfs for the pure components computed with the
parameters $\{a^l_i\}$ adjusted on the ab-initio \rdfs \textit{in the alloy},
along with experiment for pure Cu \cite{Waseda}, Zr \cite{Waseda}, and Ti
\cite{Ti-exp}.

Similarly to the ab-initio ones, the \rdfs for the pure components determined
with the parameters $\{a^l_i\}$ \textit{adjusted on the alloy} are in very
good agreement with experiment. On the contrary, with the parameters
$\{a^s_i\}$ determined from lattice constants and cohesive energy data as in
Ref. \onlinecite{Teichler1}, $g(r)$ differs significantly from experiment (see
figure (\ref{figureCuTiZrPurs}) for Cu). The situation is thus different from
that of the pure components since it was observed in Ref. \onlinecite{Alemany}
for example that tight-binding potentials give a reasonable description of the
dynamic properties of several liquid metals, in spite of having been
parametrized on the basis of solid-state data. Determining fully transferable parameters 
from the solid to the liquid and vice-versa remains thus an important task in the future.

\begin{figure}[htbp]
\centering
\includegraphics[clip,width=7.0cm]{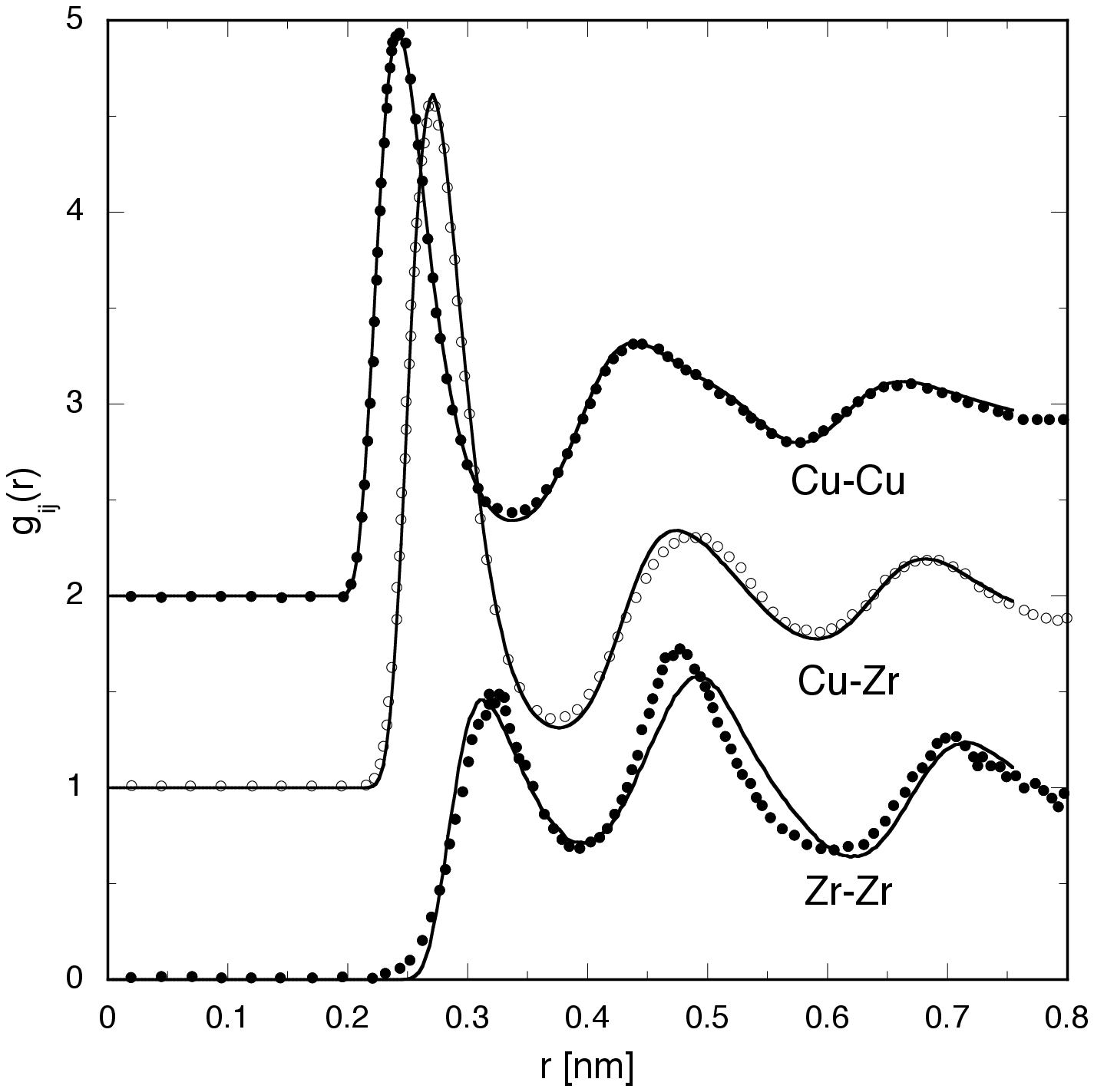}
\caption{Radial distribution functions in liquid Cu$_{80}$Zr$_{20}$ at
  $T=1500$~K.    \\Symbols: AIMD of Ref. \onlinecite{Jakse1}. Full
  lines: \SWMD, $\{a^l_i\}$.}
 \label{Cu-Zr}
\end{figure}

From the good transferability between the ternary alloy and the pure
components, one expects a similar quality of $\{a^l_i\}$ for the binary
alloys. This was checked for the Cu-Zr alloy for which data for the pair
structure are available. In figure (\ref{Cu-Zr}) we show the comparison with the
ab-initio results of Jakse and Pasturel \cite{Jakse1}. It
clearly confirms the expectation. We did not investigate other binary alloys
formed with Cu, Ti and Zr, but we see a priori no reason why the
parametrization should then behave differently.

\begin{figure}[htbp]
\centering
\includegraphics[clip,width=7.0cm]{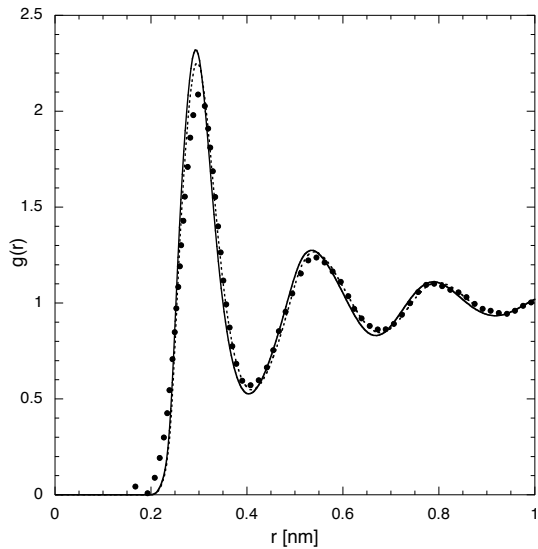}
\caption{Total radial distribution functions in the liquid Zr$_{70}$Cu$_{30}$
  alloy at $T=1453$~K.   \\Solid circles: experiment \cite{Cu-Zr_exp}; full curve:
  \SWMD, parameters $\{a^l_i\}$; dotted line: same without the volume term in
  the pressure to test sensitivity to the average density.}
  \label{gtot_Cu-Zr}
\end{figure}

A test of different nature is to compare the predicted structure directly
with experiment. We did this with the data of Ref. \onlinecite{Cu-Zr_exp} for
the Zr$_{70}$Cu$_{30}$ alloy at $T=1453$~K. This constitutes a severe test
since Cu is a minority component for this composition. $g_{tot}$ then has a
large contribution from $g_{Zr-Zr}$ and $g_{Zr-Cu}$. The result is shown in
figure (\ref{gtot_Cu-Zr}). The agreement between experiment and simulation is
quite satisfactory, especially in view of the fact that the adjustment of the
parameters for Zr was made with only 52 atoms in the ternary alloy. Our
results are comparable to those of the Monte Carlo simulation of Harvey \emph{et
al.} \cite{Harvey2} using a modified EAM formalism. Studying, possibly with
refined parameters, the amorphous structure of binary Cu-Zr alloys (like
the shoulders in the first peak of $g_{tot}$ \cite{Cu-Zr_exp2}, its sensitivity
to composition \cite{Duan}, or the effect of the cooling rate \cite{Wang}) is
left for future work.

To conclude this section, these comparisons with experiments and
other simulations show that when the adjustment is made on the structure in the
liquid state, the transferability from the alloy to the pure components and
the binary alloys is very good. Thus we have the desired starting point for
investigating whether it remains so, after a quench down to ambient
temperature.

\subsection{Amorphous alloy}
\subsubsection{Structure from classical and AIMD simulations}

\begin{figure}[htbp]
\centering
\includegraphics[clip,width=7.0cm]{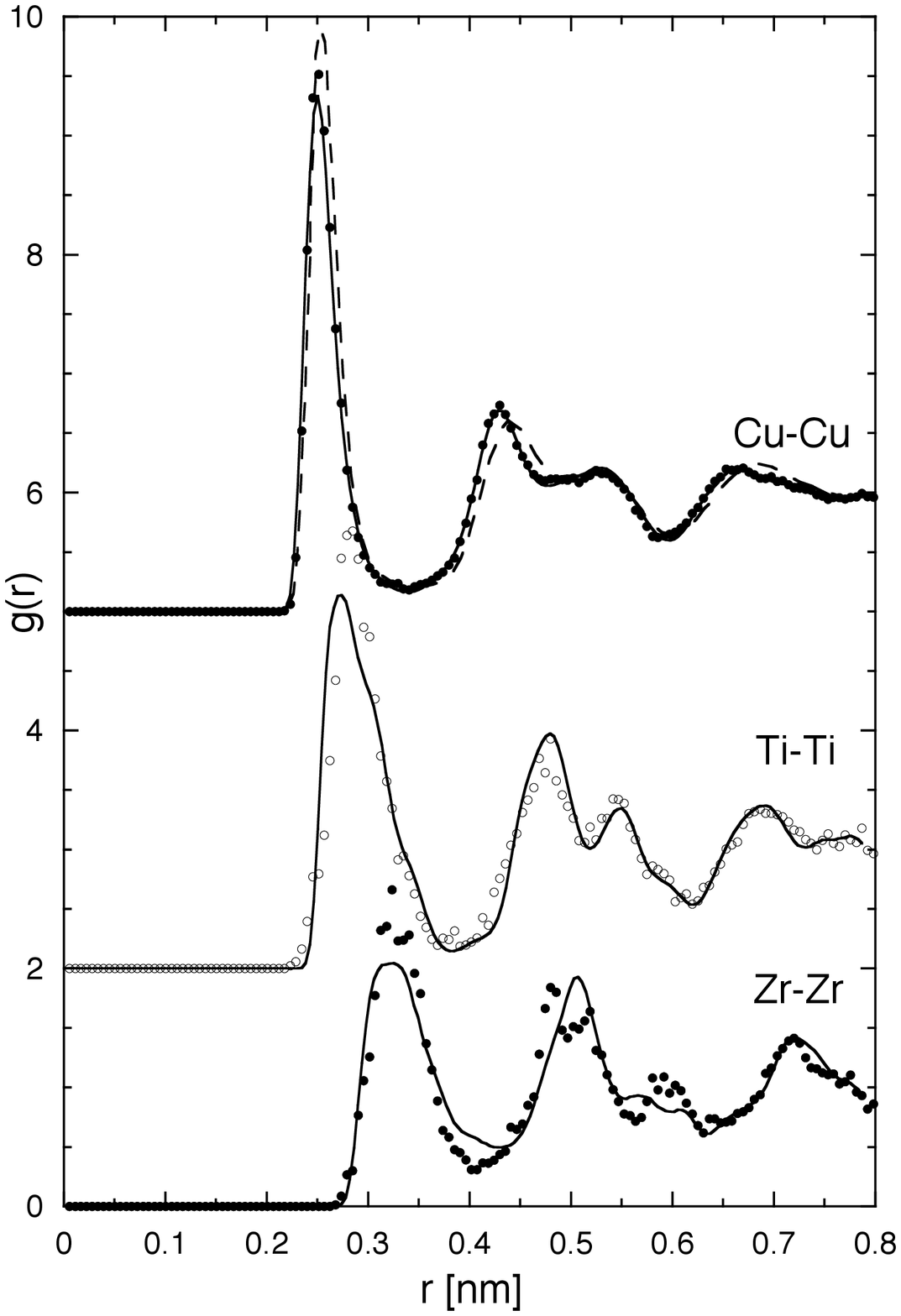}
\includegraphics[clip,width=7.0cm]{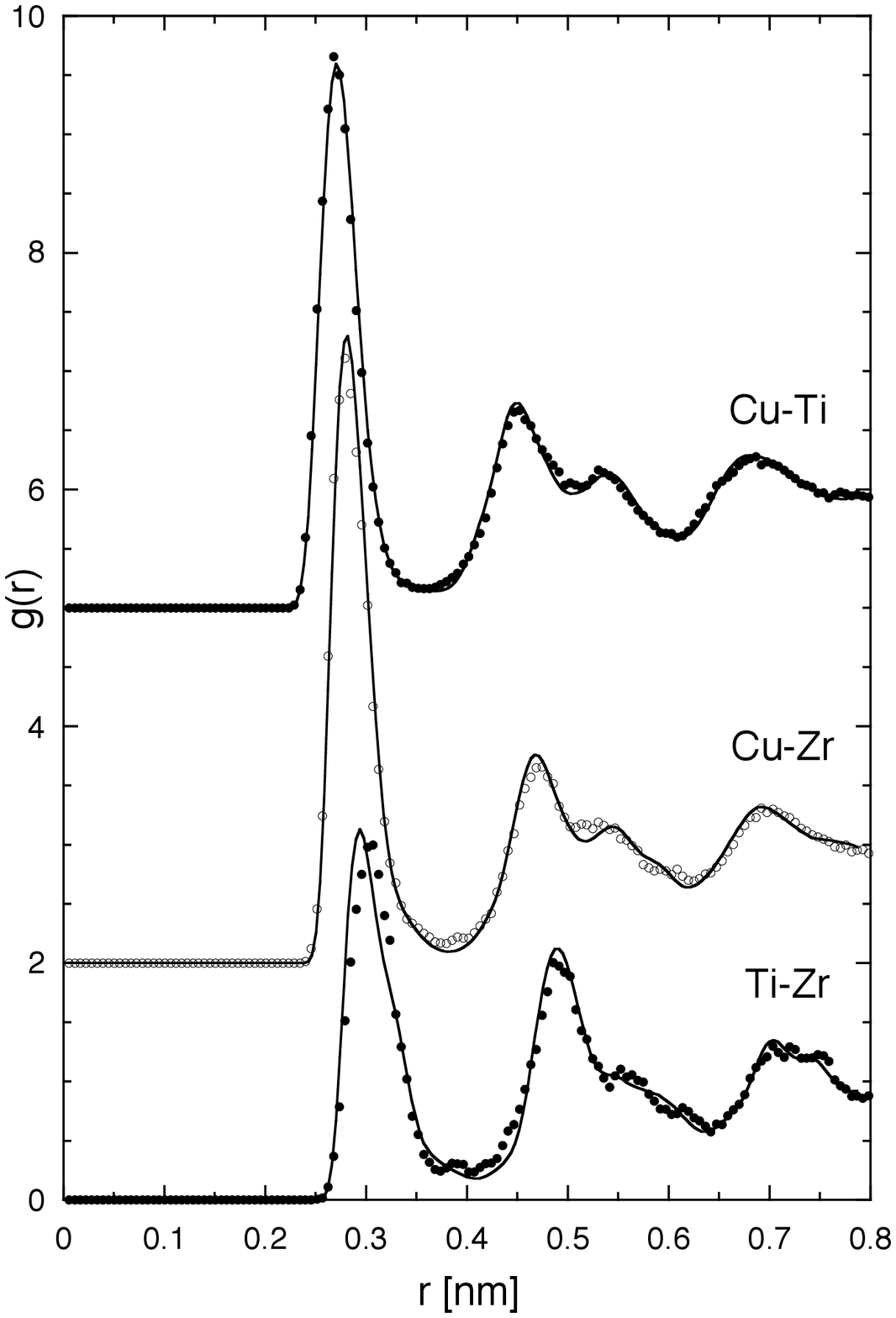}
\caption{Partial \rdfs in amorphous Cu$_{60}$Ti$_{20}$Zr$_{20}$ at 300~K.  \\
  Symbols: AIMD; full lines : \SWMD, $\{a^l_i\}$; dashed line for Cu: MD
  simulation \cite{Dalgic} with tight-binding potentials for
  Cu$_{50}$Ti$_{25}$Zr$_{25}$.}
\label{figure300K}
\end{figure}

Using the parameters $\{a^l_i\}$ in table I in the \SWMD simulation, we
perform a quench from $T=1700$~K to $T=300$~K, first at the cooling rate of
$3\,10^{10}$~K\,s$^{-1}$ used in Ref. \onlinecite{Teichler1}. The \rdfs at
the final temperature are shown in figure (\ref{figure300K}). We first observe the split second peak, typical of the
amorphous solid. The first peak is also much higher than in the liquid and
also more pronounced than the one predicted with the parameters $\{a^s_i\}$
(figure (6) in Ref. \onlinecite{Teichler1}).

To test the resulting structure at $T=300$~K, we determined the AIMD
$g^{ab}_{ij}(r)$ starting from the final configuration of the \SWMD run (below
designated as configuration (1)). To reduce CPU time, as indicated, NVT
simulation was used, for the average volume of the \SWMD run at zero pressure.
The corresponding cell size is 29.837~a.u. After runs of length similar to
the high-temperature ones (about $500+1500$ steps), sufficiently smooth curves
were obtained. We first observe in figure (\ref{figure300K}) that the \rdfs
involving Cu are very close in the ab-initio and classical MD simulations. The
larger discrepancy for Ti-Ti and Zr-Zr is mostly a consequence of the smaller
statistics in the ab-initio data, obtained with only 52 atoms. This is
confirmed by the smooth behavior of the partial \rdfs involving Cu in figure, which are averaged over a larger number of pairs.

\begin{figure}[htbp]
\centering
\includegraphics[clip,width=7.0cm]{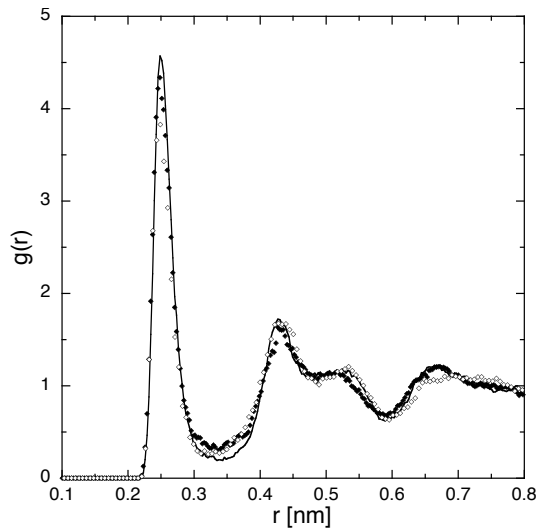}
\caption{Total ab-initio radial distribution at $T=300$~K from different initial
  configurations.\\
  Full curve: last configuration with $\{a^l_i\}$; open diamonds: last
  configuration with $\{a^s_i\}$; solid diamonds: ``instantaneous'' quench.}
 \label{test_trempe}
\end{figure}

\begin{figure}[htbp]
\centering
\includegraphics[clip,width=7.0cm]{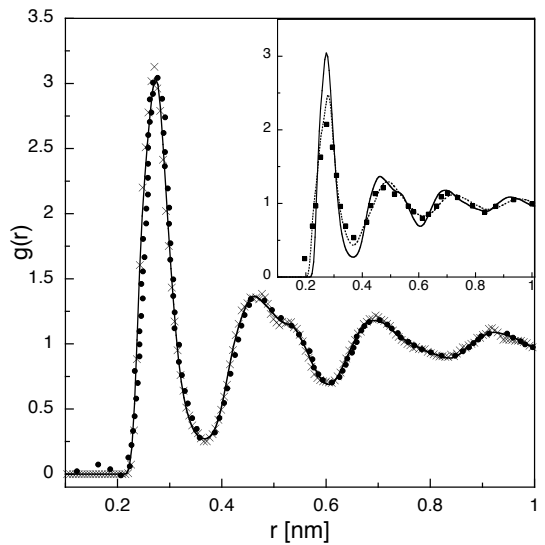}
\caption{Experimental and simulated total \rdf of amorphous \tern at
  $T=300$~K.\\
  Solid circles: XRD data of \v{D}uri\v{s}in \textit{et al.}
  \cite{Cu-Ti-Zr_Durisin}, crosses: AIMD; full curve: \SWMD.\\
%  Inset: squares: XRD data of Ref. \onlinecite{Mattern}; dotted line: \SWMD
%  with $\{a^s_i\}$, full curve: same with $\{a^l_i\}$.
}
  \label{gtot_exp}
\end{figure}

To check equilibration, the pressure was recorded during the last 750 steps.
It oscillates about a stable (albeit slightly too high) average
value\footnote{See supplemental material at \texttt{[pressure\_300\_1600.pdf]}
  for the plot of the pressure.}. This should not affect much the pair
structure, since the amorphous alloy is rather dense at $T=300$~K
($\rho=0.0661$~\AA$^{-3}$ to be compared to $\rho=0.0594$~\AA$^{-3}$ at
$T=1600$~K). Since the dynamics is particularly slow in the amorphous state,
sensitivity to the initial configuration was tested by starting the AIMD run
from two other equilibrium configurations : (2) the AIMD one at $T=1600$~K,
after 2000 steps -- this amounts to performing an instantaneous quench (the
positions were rescaled by the density ratio $(0.059/0.066)^{1/3}$); (3) the
\SWMD one at $T=300$~K, after $6\,10^6$ steps with the parameters $\{a^s_i\}$.
The latter differs from configuration (1) (the one with the parameters
$\{a^l_i\}$). The results shown in figure (\ref{test_trempe}) indicate that
(i) the initial condition plays a role, which is understandable for amorphous
states, but the final AIMD \rdfs are not strongly affected. (ii) the ab-initio
simulation is not frozen in the initial condition as evidenced by the
comparison between the final AIMD \rdfs at 300~K and those corresponding to
the initial conditions: $g^{ab}_{Cu}(r)$ is very different from that in the
liquid (configuration (2)), even for a liquid having the density of the
amorphous alloy\footnote{See supplemental material at
  \texttt{[g\_Cu\_0.066.pdf]} for $g_{Cu-Cu}(r)$ at $T=1600$~K and
  $\rho=0.066$~\AA$^{-3}$.}; it is also clearly different from that in the
amorphous state as predicted by \SWMD with parameters $\{a^s_i\}$
(configuration (2))
% ICI
-- see also inset in figure (\ref{gtot_exp}) for
$g_{tot}$. Another indication is the mean squared displacement. The AIMD value
estimated on the (limited) number of available configurations, $\Delta r^2
\sim (0.06\pm0.01)$~\AA$^2$, is compatible with the one determined by
classical MD \footnote{See supplemental material at
  \texttt{[MSD\_300K.pdf]}}, both being different from the one obtained by
classical MD with parameters $\{a^s_i\}$. This suggests that the latter  parametrization leads to  a faster dynamics in the frozen state. A larger diffusivity is also clearly evidenced by the behaviour of $\Delta r^2(t)$ in the liquid, as determined by classical simulation over a wider time range (the same should hold also in the undercooled states) \footnote{See supplemental material at
  \texttt{[MSD\_1700K.pdf]}.}.

The behaviors of the AIMD simulation with configurations (2) and (3) are then
clearly distinct from the one observed with configuration (1), that leads to
the agreement with experiment. While one might naturally fear freezing for
relatively short AIMD runs, this dependence on the initial configuration shows
that the microstates explored by the AIMD simulation remain close to those of
the classical MD only when they correspond to the correct free energy
minimum\footnote{See supplemental material for a view of the total particle
  displacements during the AIMD runs: \texttt{[displacements.png]}, run
  starting from configuration (1); \texttt{[displacements\_a\^{}s.png]}, run
  starting from configuration (3). See also the snapshot of configuration (1)
  at \texttt{[snapshot.png]}.}. This is reassuring in view of the agreement
between the classical and ab-initio \rdfs in figure (\ref{figure300K})  when using the parameters $\{a^l_i\}$.

Regardless of the connection between the structure predicted by simulation and
the one in the macroscopic alloy (see figure (\ref{gtot_exp})), one
first important result is thus the good transferability of the potentials from
the liquid to the amorphous state.

A second important observation is the noteworthy similarity of the structure
predicted for \tern to that of the Cu$_{50}$Ti$_{25}$Zr$_{25}$ alloy, with
similar composition, studied by Senturk Dalgic and Celtek \cite{Dalgic} using
TB potentials in the simulation (dashed line in figure (\ref{figure300K})).
The amorphous state of this alloy was also obtained from a quench of the
high-temperature liquid (the predicted densities at 300~K, $\rho\sim
66.13$~nm$^{-3}$ and $\rho\sim 63.09$~nm$^{-3}$, respectively, are close as
well). Such consistent predictions for similar systems deduced from different
potentials, adjusted on different data, and the fair agreement with the AIMD
for \tern constitute a strong evidence that the predicted structure is the
actual one, when the amorphous state is reached through a rapid quench from
the liquid.

This is confirmed by comparison with experiment in figure
(\ref{gtot_exp}). The agreement between simulation and the recent X-rays
diffraction data of \v{D}uri\v{s}in \textit{et al.} \cite{Cu-Ti-Zr_Durisin} is
nearly perfect.
This is quite remarkable, considering the absence of adjustable parameters in ab-initio simulations (aside from the use of pseudopotentials, which are
purely atomic properties).

\begin{figure}[htbp]
\centering
\includegraphics[clip,width=7.0cm]{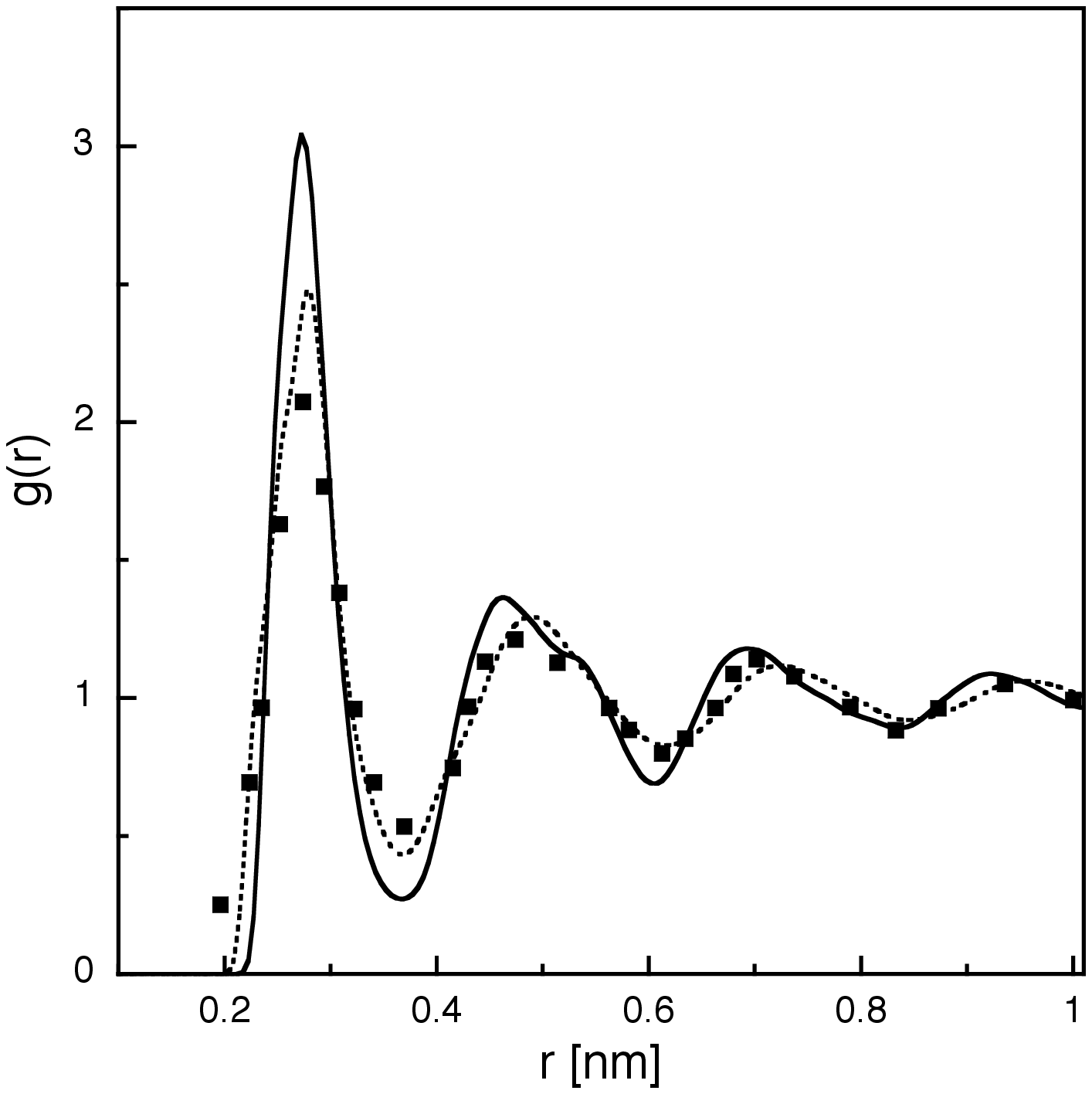}
\caption{Effect of the parametrization on the  total \rdf at
  $T=300$~K.\\
Dotted line: \SWMD with $\{a^s_i\}$, full curve: same with $\{a^l_i\}$. Squares: unpublished XRD data of Ref. \onlinecite{Mattern}  shown in Ref. \onlinecite{Teichler1}.}
  \label{gtot_Mattern}
\end{figure}

This pair structure is clearly different from the one predicted by Han and
Teichler using the $\{a^s_i\}$ parametrization of the SW potential (figure
(\ref{gtot_Mattern})). Since these parameters fitted on the ordered solid
cannot describe the structure of the liquid either, as shown above, both the
high- and low-temperature limits suggest a possible problem with this
parametrization. The puzzling point however is that the same parametrization
predicts a total \rdf $g_{tot}(r)$ in rather good agreement with the
(unpublished) experimental data \cite{Mattern} shown in Ref.
\onlinecite{Teichler1} (figure (\ref{gtot_Mattern})).

Given the excellent agreement between simulation and the experimental data of
\v{D}uri\v{s}in \textit{et al.} \cite{Cu-Ti-Zr_Durisin}, which are clearly
different from those of Ref. \onlinecite{Mattern}, and excluding artifacts, one
possibility is that the experimental \rdfs in figure (\ref{gtot_exp}) are
different because they actually correspond to different amorphous states. This
idea of some dependence on the thermodynamic path through which the amorphous
state is reached is found for example in Refs.
\onlinecite{Duan,Polyamorph_McMillan,isro,polyamorph}. The parametrization with
$\{a^s_i\}$ -- which is not validated by AIMD -- might thus
correspond to a metastable state close to the one reached in Ref.
\onlinecite{Mattern} (the path for these data \cite{Mattern} is likely the
same as in Ref. \onlinecite{Mattern2}, which does not show the \rdf). A slight
change in the thermodynamic parameters or in the parametrization should thus
drive the system away from a local free energy minimum. To test this, we first
performed a 10 times slower quench ($3\,10^{9}$~K\,s$^{-1}$), with the
parametrization $\{a^l_i\}$. We indeed find a detectable effect on
$g_{tot}(r)$. This differs from the study in Ref. \onlinecite{Wang} of the
amorphous Cu$_{46}$Zr$_{54}$ metallic glass, which showed no significant
effect of the quenching rate in the higher range investigated. The authors
used TB-SMA n-body potentials parametrized according to the method of Ref.
\onlinecite{Duan}, in which the force-field parameters were obtained from a
fit to first-principles/DFT calculation, as we do it here. The slight effect
of the cooling rate we found is however insufficient to explain the difference
between the two experiments. A path dependence is also suggested by the solid
solution model in Ref. \onlinecite{Qin}. It is difficult to compare our
results with those for the Cu$_{70}$Zr$_x$Ti$_{1-x}$ systems, but the great
sensitivity to the concentration of the minority species shown clearly
testifies for the importance of the path leading to the amorphization -- the
solid solution remaining crystalline at some concentrations, for example (see
the discussion of figure (8) in Ref. \onlinecite{Qin}).

In the numerical quench experiment, another source for the observed
discrepancy is the initial structure, which, as discussed above, is quite
different with $\{a^s_i\}$ and $\{a^l_i\}$. As a test of the sensitivity to
the parametrization, we relaxed the adjustment of the strength parameter for
Cu to start with a structure that is in-between the ``correct'' one (i.e. with
$\{a^l_i\}$) and the one obtained with $\{a^s_i\}$ (recall that the latter
predicts a less structured liquid). As expected, the resulting $g_{tot}(r)$ at
300~K (not shown here) is indeed in between the one obtained with the initial
$\{a^l_i\}$ parameters (or experiment) and the one with $\{a^s_i\}$. Another
indication suggesting metastability of the state described by the data of Ref.
\onlinecite{Mattern} is the fact that the AIMD run quickly departs from
configuration (3) (see also figure (\ref{test_trempe})). More experiments
would thus be useful to ascertain the structure in amorphous states reached
through different thermodynamic paths.

\subsubsection{Glass transition temperature}
\begin{figure}[htbp]
\centering
\includegraphics[clip,width=7.0cm]{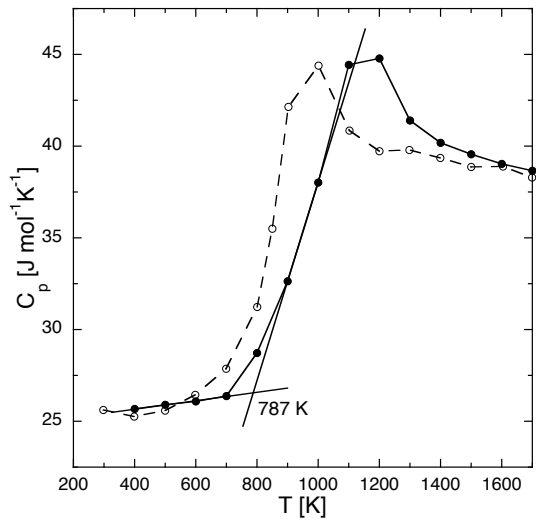}
\caption{Heat capacity versus temperature.\\
Full curve: \SWMD with $\{a^l_i\}$; dashed curve: same with $\{a^s_i\}$.}
  \label{Cp-T}
\end{figure}

The simulation of a small periodic systems is not expected to give
an accurate description of the amorphization of the real ternary alloy,
besides other questions such as surface effects in the experiment. Its
computational cost for the required larger systems discouraged us from
attempting it in this work. Nevertheless, to test the idea of adjusting the
parameters in the liquid, we used the classical \SWMD route with the
parameters $\{a^l_i\}$ to estimate the critical temperature of the glass
transition. This was done from the behavior of the constant-pressure heat
capacity $C_p$, shown in figure (\ref{Cp-T}).

We thus find $T_g\sim787$~K, slightly above the value determined with the
parameters $\{a^s_i\}$ ($T_g\sim750$~K).  This is consistent with the  diffusivity being lower with the parameters $\{a^l_i\}$. Both are above experimental estimates
($T^{exp}_g\sim710$~K in Ref. \onlinecite{Inoue2} or $T^{exp}_g\sim695$~K in
Ref. \onlinecite{Cu-Ti-Zr_Durisin}). Comparison with experiment is however not
immediate for several reasons, including the cooling rates that are achievable
in simulation \cite{Teichler1}.  The higher cooling rates in the simulation should indeed overestimate the critical temperature.  A clarification of this aspect on the basis of
ab-initio simulation is left for future work.

\subsection{Finite-size effects}\label{sectionDOS}
The main question we investigated here is the transferability at different
temperatures and compositions of parametrized potentials for a representative
ternary alloy, using, for computational convenience, a rather small system.
This question is actually independent of the closeness of the properties of
the small periodic system to those of the macroscopic alloy. Aside from the
question of statistics, using a small system does not indeed constitute a
limitation as long as the goal remains to test transferability. To have an
idea of the influence of the system size, some calculations were nevertheless
repeated using $N=240$ atoms, and no significant change of the pair structure
was found.

Actually, the possible impact of using a small periodic system has two
different aspects: the first one concerns the variation of the structure with
the number of atoms considered. The second one is the metallic character of a
small system, especially when the minority species are represented by a few
tens of atoms.

The first aspect can be important due to the lack of long-range order of
amorphous materials which requires using blocks containing at least several
hundreds of particles (see Ref. \onlinecite{Kulp} for Cu-Ti). To investigate
this, we repeated the classical simulations with much larger systems. As we
found no significant effect on the \rdf, the artificial periodicity related to
a small system size should not be a qualitative limitation, as far as the
average structure is concerned. This is confirmed by the very good agreement
with the recent X-ray data shown in figure (\ref{gtot_exp}). On the other
hand, the ab-initio \rdfs were obtained in NVT conditions for a rather
small system having the average density at zero pressure in the classical MD.
This gives a slightly too high ab-initio pressure while the small system size
makes it slightly anisotropic. But due to the very steep variation of the
pressure with density only very accurate values for the latter might require
larger systems.

Other structural properties might however be much more sensitive to the size
of the system (see e.g. Refs. \onlinecite{Cheng2,Cluster-Glue2,Fujita}). In
Ref. \onlinecite{Fujita}, for example, it has been noticed that selective
minor additions can dramatically improve the GFA of binary metallic glasses.
In particular, the effect of composition on the glass formation of the
Cu-Ti-Zr alloys has been discussed in Refs. \onlinecite{Dai,Dai2,Ze-xiu,Qin}.
For the family of Cu-based alloys, a recent study \cite{Ward} based on the
parametrization of Ref. \onlinecite{Mendelev} evidenced the subtle structural
evolution that occurs during cooling. They pointed out that the presence of
icosahedrally coordinated clusters and their tendency to form networks is
insufficient to explain glass formation at all compositions in the Cu-Zr
binary system.
Besides the Cu-Ti-Zr alloy considered here, there are several studies of
Cu-based ternary alloys, such as Cu-Zr-Al \cite{Cu-Zr-Al} and Cu-Zr-Ag
\cite{Cu-Zr-Ag}. In Ref. \onlinecite{Cu-Zr-Ag} a many-body potential was
developed using the embedded atom method (EAM) on the basis of ab-initio
calculations \cite{Cheng2}. They pointed out the coupling between chemical and
dynamical heterogeneities, which appears to play a crucial role in the
improved GFA of this alloy and the Cu family of alloys studied in Ref.
\onlinecite{Cu*}. Studying such size-dependent structural effects by ab-initio
simulations would likely require larger samples than those considered here.

\begin{figure}[htbp]
\centering
\includegraphics[clip,width=7.0cm]{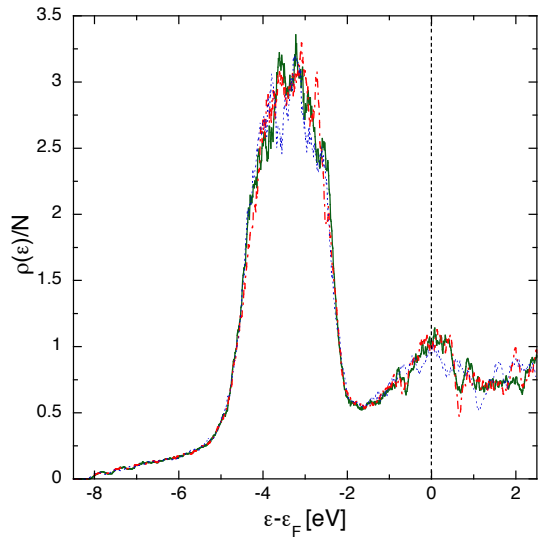}
\caption{(color online) Instantaneous total DOS in typical configurations at
  300~K. Dotted line: $N=260$; full and dash-dotted lines: $N=240$.
  Calculations were made with 36 k-points in the tetrahedron
  method \cite{tetrahedron}.}
 \label{3Dos}
\end{figure}

The second question relates to how far the electronic structure would have
been different had we used a larger system, due to the actual sensitivity of
the metallic character to the number of atoms in a small sample, as discussed
in the literature on the related field of metallic clusters (see e.g. Refs.
\onlinecite{Cluster1,Cluster2,Cluster3} and references therein). For the
actual size used here, an idea can be formed from the system size dependence
of typical electronic properties such as the density of states (DOS), the
Fermi level, or the energy per particle. Figure (\ref{3Dos}) shows the
instantaneous DOS in equilibrated configurations of the nuclei at 300~K: two
for $N=240$ atoms and one for $N=260$. In the three cases, the DOSs are very
similar, being dominated by the d-band-like contribution of Cu in the range
$-6~\text{eV}\leqslant \epsilon-\epsilon_F \leqslant -2$~eV. The corresponding
Fermi energies are $\epsilon_F=13.4181$, $13.4228$, and $13.3775$~eV, and the
total energies per particle are $E/N=-1578.922$, $-1578.919$, and
$-1578.887$~eV, respectively. A quite similar behavior for Cu-Zr binaries is
shown in figure (4) of Ref. \onlinecite{DOS_Cu-Zr}. This shows that the
variation with system size is of the same amplitude as the variation between
two configurations having the same number of particles. Averaging over a large
number of configurations will smooth even more the tiny differences found here
(see also Ref. \onlinecite{DOS_Cu-Zr}). Thus the systems we used seem large
enough also for the electronic structure at fixed configuration of the nuclei.

\section{Conclusion}

We have discussed in this work the transferability of a simplified model
potential of the Stillinger-Weber form, between the pure components and the
alloy in the liquid state, and for the alloy during a quench down to the
amorphous state. Comparison with ab-initio simulation shows that the
parameters adjusted in the liquid are transferable at high temperature
independently of the composition, and well below the critical glass
transition, as far as the pair structure is concerned. This contrasts with the
parameters adjusted in the solid which cannot describe the structure in the
liquid, and predict an amorphous structure that is inconsistent with
ab-initio simulation and the more recent diffraction experiments. This rises
the question of the path followed to reach possibly metastable amorphous
states, prior to the comparison with theoretical predictions. 
Concerning the question of the size of the simulated system, it is found from
a discussion of the size dependence of the average structure, and of the
metallicity that ab-initio simulations are feasible on medium size computers
while the study of other properties such as local structural organization, of
low cooling rates, or simulations in NPT conditions still require larger
computational resources. Comparison with the prediction of the critical
temperature of the glass transition and the available experimental structure
using classical simulation, which is less subject to system size limitations,
finally underlines the importance of developing thoroughly tested parametrized
force fields, in order to keep simulations convenient enough for a
quantitative study of the behavior of complex materials.

\end{document}